# A Customised App to Attract Female Teenagers to Coding

Bernadette Spieler, Wolfgang Slany
Graz University of Technology, Graz, Austria
bernadette.spieler@ist.tugraz.at
wolfgang.slany@tugraz.at

**Abstract:** The number of women in IT-related disciplines is far below the number of men, especially in developed countries. Middle-school girls appear to be engaged in coding courses, but when they choose academic majors relevant to their future careers, only few pursue computer science as a major. In order to show students a new way of learning and to engage them with coding activities, we used the learning app Pocket Code. In the "No One Left Behind" H2020 European project, the app was evaluated in several school subjects. An evaluation of the attractiveness of the app shows that students were motivated by Pocket Code's ease of use and its appealing design; however, girls rated the app less enthusiastically. To appeal to female teenagers in particular, a tailored version of the app "Luna&Cat" has been developed. This customised version stands in contrast to the "one size fits all" solution Pocket Code, which may discourage certain user groups. For apps to have a higher chance to appeal to a specific target group, it is, among many other points, necessary to optimise their store listing on app stores, especially as we found that app stores are the most effective way to reach teenagers. Thus, this paper covers the following research question: What customizations are necessary in Pocket Code to reinforce female teenagers in their coding activities? To answer this question, a focus group discussion was performed. This discussion first brought insights about our target group and suggested names and designs for the new app; and second, allowed each student to make proposals for their desired games. Later, these game ideas were analysed, graphically designed, and further developed together with university design students. By showing female teenagers games designed by other young women in their age group, we help them to get ideas and inspiration to code their own programs. This is important because most girls have the feeling that the games they play are not created for them. With this customised app, our aim is to reach and build a user base of interested female teenagers who want to learn how to code.

## 1. Introduction

The literature is full of findings that document girls' low experience or affinity levels towards computer sciences, their negative attitudes, and their fear to fail in computer science subjects compared to their male colleagues (Zaidi et al., 2016; Ramos and Rojas-Rajs, 2016; Wong and Kemp, 2017). Thus, the research focuses on female deficits in programming. However, what if these are not deficits but just different approaches or preferences that coding tools of today do not allow or fulfil? By understanding these differences, we hope to support girls so that they become more interested in coding. In previous work the authors focussed on how to create inclusive learning environments for girls (Spieler, 2018; Spieler and Slany [2], 2018), and analysing gender differences in game design (Spieler and Slany [1], 2018). To support our female user group outside of the school setting, a new flavored version of Pocket Code has been developed with the name "Luna&Cat. Its goal is to empower our female users while playing games through new game design possibilities; this is done with engagement through gender sensibility and awareness in contents used, creativity through allowing personalization and customization, and coding by problem-based tasks to foster self-directed learning and self-expression. Concerning Pocket Code, it is uncertain if female teenagers who search for such a tool will find or download it. We see a much greater chance that the newly customised version Luna&Cat will be more appealing for girls.

This paper is organised as follows: First, in Section 2, the paper provides an overview about coding, games, and gender issues in tools. This should provide evidence for a customised version for girls and suggests customization for the new version. Subsequently, Section 3 presents the base app, Pocket Code, and the related "No One Left Behind" (NOLB) project. Section 4 shows the Hassenzahl model used for analysing the attractiveness of our app. Section 5 displays the results by gender and Section 6 discusses the results. Section 7 presents a first concept of the new Pocket Code flavor Luna&Cat, and Section 8 concludes this paper and describes the authors' future work.

## 2. Literature Review

The acquisition of digital skills is more important than ever and represents a key professional qualification. There is a great potential for young women to counteract the acute (and growing) shortage of qualified professionals in ICT (European Commission, 2016). However, the absence of female students who are interested in ICT fields can be observed at all levels of education as well as in the industry (Lamborelle and Fernandez, 2016). To allow women to have better career choices, to get quality jobs, and even to improve their lives, opportunities must be visible. To provide a clearer image of the issues that have emerged, this section summarises CS issues in schools, and concludes by presenting opportunities to learn coding outside schools through tailored tools.

### 2.1 Acquiring Computational Thinking Skills.

There are many efforts in Europe and worldwide to foster students, particularly female teenagers, in CS (Informatics Europe/ACM Europe, 2016). However, strategies to redesign and rework the curriculum are obtained slowly, current solutions are often not applicable, and possibilities are limited. Thus, education programs neither offer methods nor framework for CS classes, nor get teachers trained specifically for this subject. Consequently, many teenagers leave school without any meaningful knowledge in CS, never quite understanding what CS is and how it relates to algorithmic thinking or problem-solving (Giannakos et al., 2014). These students will be less likely to choose a career in CS or study it as a major. Additionally, across Europe there are a number of extracurricular initiatives, e.g., Educational Robotics, Makerspaces, or FabLabs. However, these coding initiatives have predominantly male participation (Zagami et al., 2015). To promote initiatives for female teenagers as girls-only, e.g., coding clubs or summer camps are a good opportunity (Alvarado, Cao, and Minnes, 2017). On the one hand, such initiatives serve as vehicles to interest girls more deeply in CS and to foster their sense of belonging and self-efficacy. On the other hand, such initiatives exclude female teenagers who live outside the city (often such activities are promoted in bigger cities), girls that are underprivileged (e.g., initiatives have fees and additional costs), and they are quickly fully booked.

One accessible way to reach out to the remaining group of female teenagers that has no opportunity to learn coding inside or outside school, is to catch their attention with attractive, supportive tools and games. Games and Game Based Learning (GBL)-tools are effective approaches to motivate teenagers to interact and to learn (Kafai and Vasudevan, 2015). Statistics show that mobile gamers are more likely to be female, have a higher income, and are younger, compared to the online population (Verto Analytics, 2015). In addition, there is already a small shift in putting more focus on gender inequality issues in games (Jenson et al., 2007; Williams et al., 2009; Martin and Rafalow, 2015). However, most developed games only appeal to a male-dominated audience and exclude female gamers; there are less successful games available for girls than there are for boys (Google Inc., 2018), most games do not appeal to a female audience, and there exists less motivation for girls to become gamers (NewZoo, 1017). Moreover, video games with female protagonists are in the minority (Jayanth, 2014). It is important to know that there exists a correlation between students who play video computer games and those who rate their computer skills as good (Davies et al., 2014). To conclude, building gender-appropriate and gender-sensitive games and GBL tools are excellent ways to attract female teenagers to CS.

### 2.2 Customised services for female teenagers.

Some may argue against developing tools/products specifically designed for women. On the one hand, many examples exist where companies are marketing to women as a specific target group with unique needs and preferences, e.g., clothes, or the color pink (Atlanta Parent Editorial, 2017; Rommes et al., 2011). Certain IT products that would seem to be clearly unisex, e.g., laptops, printers, or business software, are nevertheless not undisputed (e.g., "Petticoat 5"[1] from the 80s, "Floral Kiss" from today). Such products are so called "pink technologies" created to appeal to female customers (Rommes et al., 2011). In other cases, it is less controversial to take into account the needs or interests of women, e.g., in speech recognition systems, in games (Google, 2018), or inclusive educational software (e.g., Girl Games[2], Purple Moon (Gorriz and Medina, 2000)). According to a white paper by (Google Inc., 2018), to create inclusive games a developer should 1) know the audience (current and potential), and 2) look at game designs that may exclude potential players (imagery, iconography, etc.), and test it by store listing experiments. A study by (Martinson, 2005) analyzed the influence of having a diverse developer team (range of different perspectives). In addition, for Google Play it is important to pay attention to the store icon, screenshots, videos, and launching of female characters (see Section 7). Another discussion is about inclusion of gender perspectives in programming environments and software (Huff, 2002, Burnett et al., 2010) or feature discovery (Beckwith et al., 2006). Researchers argue that most environments are designed by men, and thus women have to adopt a male perspective when working with these technologies. Male-dominated environments have a negative impact on women's choices and reinforce gender stereotypes. Thus, an inclusion of gender perspectives during the planning and development phase is important, as is considering the impact of gender relationships for product design.

---

[1] Petticoat: https://www.chonday.com/15509/marketing-computers-for-women-in-the-80s/
[2] Girls Games: https://www.wired.com/1997/04/es-girlgames/

### 3. The Pocket Code App and the "No One Left Behind" project

The No One Left Behind[3] (NOLB) project started in January 2015, was funded in the Horizon 2020 framework of the European Commission and was conclude in June 2017. The goal of NOLB was to unlock inclusive gaming creation/game design and to construct experiences in formal learning situations, particularly for teenagers at risk of social exclusion. The study in Austria was seen as a chance to recognize gender differences in engaging with coding and game design activities (Beltràn et al., 2015, Petri et al., 2016). Thus, the app Pocket Code[4], a media-rich programming environment for teenagers to learn coding with a visual programming language very similar to the Scratch[5] environment, has been used for creating games and apps (Slany, 2014; Slany et al., 2018). Pocket Code is freely available on the Google Play Store and allows the creation of games, stories, and many types of other apps directly on phones, thereby teaching fundamental programming skills. In addition, a new variant of the app with the name Create@School was developed. It uses the same UI as Pocket Code but comes with predefined game templates for pupils and corresponding instructions for teachers

### 4. Method

During the second cycle of NOLB, an AttrakDiff survey[6] was conducted with the Create@School app. The collected data from the AttrakDiff survey is evaluated through a Hassenzahl model (Law et al., 2017). This model evaluates (1) the usability and utility perceived by users, (2) the satisfaction of users with the app, and (3) the attractiveness of the app. Based on the Hassenzahl model, the qualities of a software are classified into two distinct groups:
- *Pragmatic qualities*: they relate to practicality and functionality: the purpose of the software should be clear and understandable for the user.
- *Hedonic qualities*: they reflect the psychological needs and emotional experiences of the user. Hedonic qualities are divided into two categories:
  - *Stimulation* (HQ-S): The user wants to be stimulated in order to enjoy the experience.
  - *Identification* (HQ-I): Users have a need for expressing themselves through objects and have a desire to communicate their identity to others.

### 5. Results

For the evaluation, pupils from Austrian and Spanish pilot schools filled out the survey (152 girls, 198 boys). The results are presented below (Spieler and Mashkina, 2017).

---

[3] No One Left Behind EC project H2020: http://no1leftbehind.eu/
[4] Pocket Code Google Play Store: https://catrob.at/pc
[5] Scratch MIT: https://scratch.mit.edu/
[6] AttrakDiff-survey: http: //attrakdiff.de/

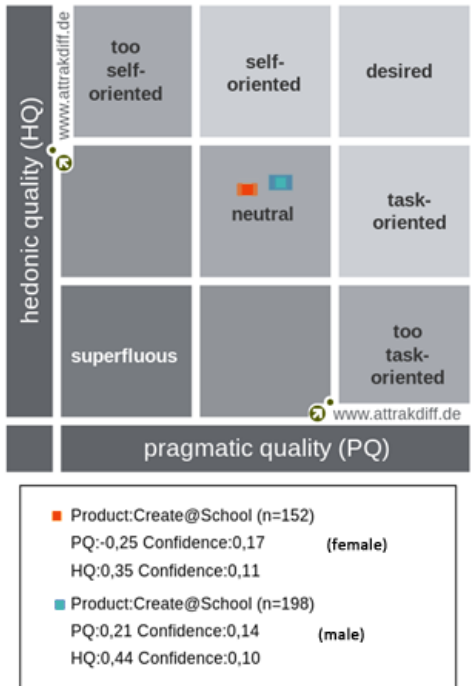 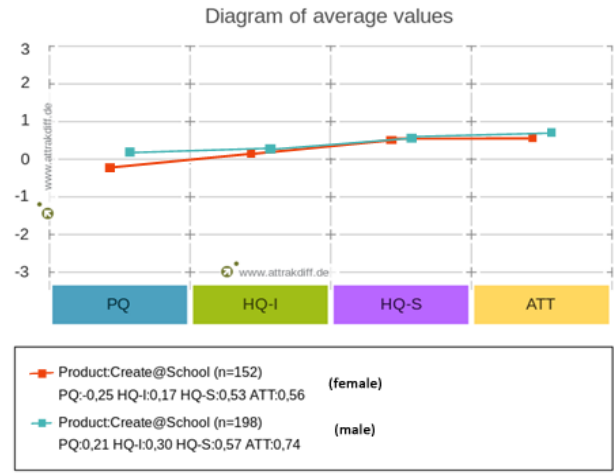

Figure 1. Overall evaluation of hedonic and pragmatic qualities.

Figure 2. Diagram of the average values for pragmatic qualities (PQ), hedonic qualities — identity (HQ-I), hedonic quality stimulation (HQ-S), and attractiveness (ATT).

As can be seen in Figure 1, Create@School received neutral evaluations for pragmatic and hedonic perspectives from both genders. The small confidence intervals signify that the results are reliable and not coincidental. Male pupils (cyan) rated the app as slightly more attractive than female pupils (orange) did (see Figure 2). The average value of the pragmatic quality (PQ) evaluation has the largest absolute distance of 0.46 between female (orange) and male pupils (cyan). The average value of the PQ rated by the female participants has the value -0.25 and thereby it is below 0. The identity (HQ-I) and attractiveness (ATT) average value is slightly larger for the male participants with the difference of 0.13 and 0.18 respectively. The stimulation (HQ-S) has approximately the same average values with 0.53 from female and 0.57 from male pupils. The average values of the hedonic qualities do not reach maximum values in any of the aspects.

The average diagram shows that:
- almost all qualities achieve a positive evaluation in both genders,
- the largest distance between genders is seen in pragmatic quality,
- female pupils' evaluation of pragmatic qualities is less than zero,
- for identity (HQ-I) male pupils rate Create@School slightly better than female pupils,
- for stimulation (HQ-S) female and male pupils' rates are almost the same,
- for attractiveness (ATT) male pupils rated Create@School slightly better, and
- no quality reaches the maximum rating.

The diagram of Figure 3 shows the details of the semantic differential of the model, the adjectives of the survey, and the values that have contributed to each quality. Overall, none of Create@School's qualities reach the maximum value. The negative feedback is mostly given by female pupils. Only the "cheap-premium" pair has been rated more negatively by male pupils. Female participants considered the app more technical than human, more complicated than simple, and more confusing than clearly structured. Male participants also rated the app more technical than human, but at the same time found it more practical than impractical, more straightforward than cumbersome, and more structured than confusing. Both groups had a neutral opinion on the predictability of the app. The identity (HQ-I) as well as the simulation (HW-S) evaluation of the app are mostly consistent between the two groups. Male pupils found Create@School more connective than female pupils did. Girls said that the app separates them from people rather than brings them closer to them. Boys rated the app as more bold and captivating than girls did, and also slightly undemanding rather than challenging, whereas female participants saw the app as challenging. Both groups rated the attractiveness (ATT) of

Create@School positively. The boys rated the app as more likeable, inviting, appealing, and motivating than girls did.

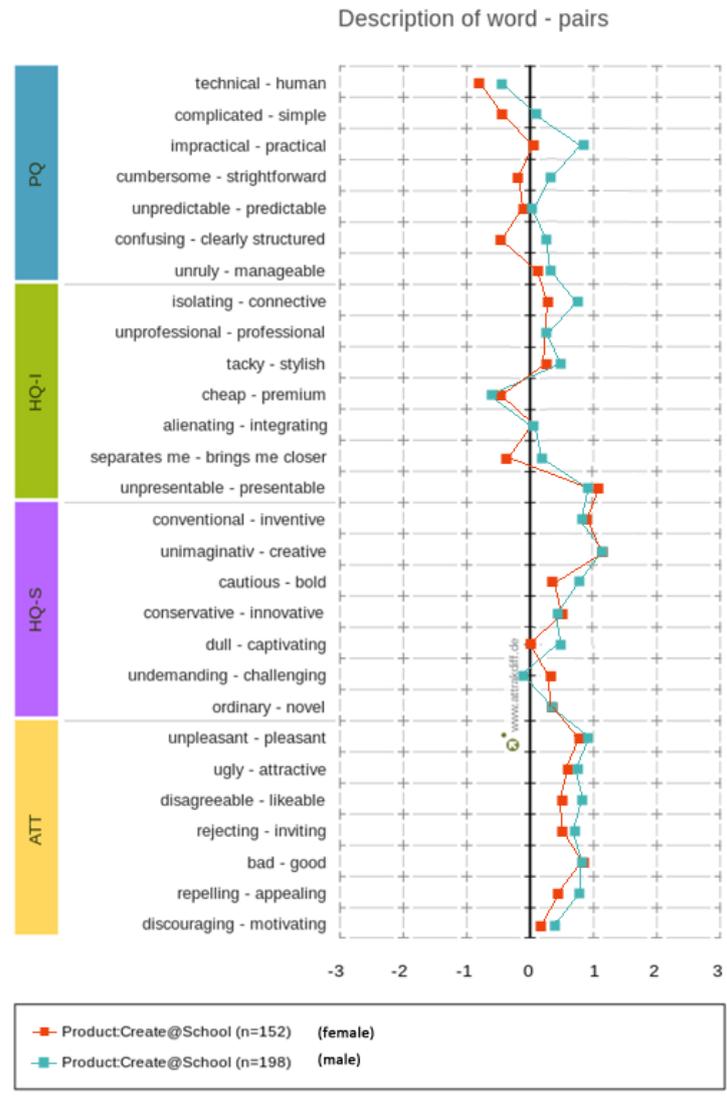

Figure 3: AttracDiff-survey: description of word-pairs.

## 6. Discussion

Create@School received very similar evaluations from both genders. In general, all pupils accepted the app and ranked it positively, even though it has not yet achieved the desired acceptance level. The positive values recorded and the good balance between pragmatic and hedonic qualities shows a degree of satisfaction while working with the application. These results of PQ are not surprising and correlate with the literature which states that there are differences between the perceptions of tools between the genders. The especially positive ratings for creativity and presentation how that participants enjoyed their experience with Create@School in general, and that the app allows users to express themselves. Regarding the negative rating for "it separates people" rather than "brings them closer", further research needs to be done if this somehow is an indicator and influences female pupils' sense of belonging or self-efficacy negatively. To conclude, the Create@School app was perceived as useful and usable, attractive and stimulating, and as being a good first step towards the creation of a new innovative tool for education that promotes the social inclusion of pupils.

# 7. Luna&Cat: a first concept to reinforce female teenagers.

The concept of "strategic essentialism" (Spivak, 1990) shows that is sometimes necessary to strengthen a specific stereotype in order to reinforce a group in the long term. During NOLB, the team got a clearer picture of what kind of gaming elements girls tend to like, create, and play (Spieler and Slany, 2018). These were obviously not the same preferences as those of boys. Thus, we decided to create a special version tailored for female teenagers in order to study the ensuing effects.

To create inclusive games and to include our audience, focus group discussions were conducted with 10 girls of two different age groups (12-13 years old, 14-15 years old). Girls talked about different names, designs, and developed game ideas with the help of a playful design activities. Different materials were prepared: questionnaires, potential app names, and idea cards for the game idea development process. Although 4 out of 10 girls said that they never play any computer games at all, all of them were convinced that they knew exactly what the game that would be most appealing to them would look like.

Subsequently, 14 university students from the first year of the master's degree program "Industrial Design" from the University of Applied Science (FH JOANNEUM) developed those ideas further as part of a 14-hour design workshop. During this workshop, 5 out of 10 game ideas from the focus group discussion were developed further. Moreover, personas from the questionnaires were created to find out more about our target group's characteristics and aims (Cooper, 2003). As a result of the workshop, mock-ups, prototypes, and finished Pocket Code programs were submitted, which later came to serve as featured games for the new app flavor. In addition, a new promo video has been created[7].

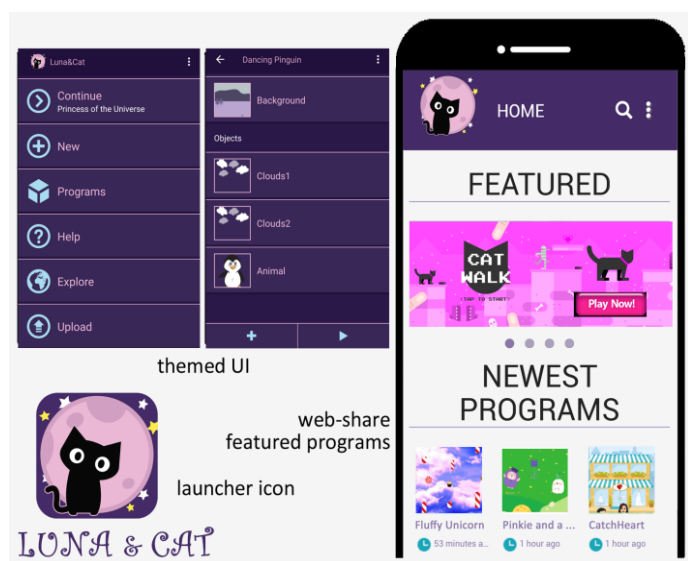

Figure 4: Luna&Cat logo, UI, and community webshare

Figure 4 shows the new app UI including the launcher icon, the themed UI, and the new community webshare view with featured games. It was essential to choose an appropriate design: The app icon plays an important role when users select apps from search results and category lists. Furthermore, screenshots and names are important, as users will see them when they select the app from the list of results. The name chosen for the new version was Luna&Cat. Luna&Cat sounds pleasant, has a connection to our project's name (Catrobat) and with Scratch, and seems to be very appealing to our focus group. The palette of violet colors used in the app was rated very positively by the focus group and had already been used successfully in a previous Alice Game Jam event[8] we conducted. The second screenshot shows the new example game "penguin dance" (the penguin was the most used character from our extensive gallery in girls' games). The new Luna&Cat app will be available at Google Play by the end of 2018. It has been developed in relation to four components recognized as important for girls (see PECC model Spieler, 2018): *Playing, Engagement, Creativity,* and *Coding*.

*Playing & Engagement*: Designing gender-sensitive games is a promising approach to close the gender gap (see Literature review). The study by (Heeter et al., 2000) found that games designed by girls were ranked higher by female teenagers than by male teenagers. In addition, they considered them to be significantly better for learning than boys' games. During the NOLB units, there were several times in class that male pupils shouted "Cool, there is a Star Wars/Alien shooter game" etc., whereas their female colleagues usually only played games created by their female peers. Currently, recommended games on the sharing site are very similar, except for the random and new programs category. For newly uploaded programs, it is very difficult to reach the most downloaded/most viewed section at all. On average, each program is downloaded 26 times[9]. Thus, we see that our users love to download and play existing games, create and design similar games, or remix existing ones (i.e., a game created on the basis of another game). Sharing is critical: Many successful apps have a very active community, as evidenced by the successes of Pinterest, YouTube, and Instagram. The community is essential to feeling engaged. People who are part of this community want to feel a sense of belonging as well. With Luna&Cat we can give girls appealing featured games created on the basis of ideas from our focus group[10] (see banners Figure 5). Some of these games provide a scene with additional graphics to add a second level.

---

[7] New promo video for Luna&Cat: https://youtu.be/-6AEZrSbOMg
[8] Alice Game Jam Event: http://www.alicegamejam.com/

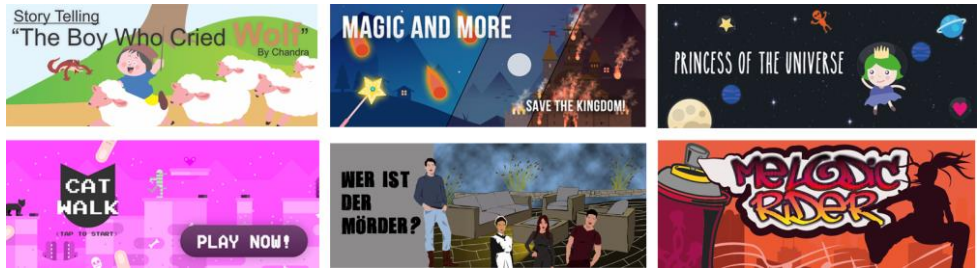

Figure 5: Banner for the new featured games of Luna&Cat.

*Creativity*: To support female pupils' creativity, we saw during NOLB that it is not only important to allow them to draw assets or to give them enough time during game design, but also to provide assets they can use, adopt, or even customise. This allows personalization through self-expression, and builds a sense of ownership. In addition, with semi-professionally designed assets the games become more beautiful and appealing.

*Coding*: By using the core functionality of the app - the coding - girls can create games and animations with the use of a visual block-based language. The authors' research showed that no differences could be observed regarding the functional structure of pupils' games based on gender (see Spieler and Slany, 2018 [2]). Comments collected by (Spieler and Slany, 2018 [1]) concerning the app showed that all users wished that the app would be more clearly structured and easier to use. Therefore, the whole team is working at the time of writing this paper on refactoring/redesigning tasks to make the app more appealing for all users. Luna&Cat has the same programming functionality as Pocket Code. To support our female pupils in their coding, missing functionalities mentioned in surveys and focus group discussions by girls during NOLB need our further consideration (e.g., media or predefined objects).

## 8. Conclusion and Future Work

With Luna&Cat, we hope that we can reach and build a new large user base of curious female teenagers who want to learn how to code. First, the featured games and a new example game should encourage female pupils and awaken their interests for coding. With this, the authors hope that in the future more games from female users will occupy the different highly visible sections on the sharing site. Second, Luna&Cat will help to support pupils in their engagement, building a new growing community of female teenagers who will love to share their games and assets with others, thus generating a sense of belonging to a community of female coders. The possibility of liking, commenting, remixing, and downloading the games will cultivate a new, engaged female community from all over the world. Third, to spark creativity, new assets and a continuously growing media library will help to support female teenagers in creating games that appeal to them. Thus, they can become proud of their programs and encouraged to share them with the community. Finally, coding will foster their self-efficacy, and Luna&Cat will be a tool for efficiently creating games.

To engage girls in coding, a new project started in September 2018, with the name "Code'n'Stitch". During this project we will extent the Luna&Cat app with the option to program embroidery machines (very similar to the existing Turtlestitch project[11]). In this way, self-created patterns and designs can be stitched on t-shirts, pants, or even bags. With Luna&Cat, the embroidery machines will get programmable. Patterns and different forms can be created using our visual programming language Catrobat. As a result, teenagers have something they can be proud of, something they can wear, and they can show to others. A special emphasis will be given to a gender-equitable conception to consider different requirements, needs and interests of our target group. These courses with schools will be realized together with partners "bits4kids[12]". On the one hand, with this option the team want to show young women new ways of using technology, with a lot of fun in a sustainable way. On the other hand, young men can get inspired too through this digital design process and the possibility of new challenges in textile handicraft lessons. A cooperation with the fashion shop "Apflbutzn[13]" will help us to picture the whole workflow. The Apflbutzn team will take part in the last units and bring their embroidery machine to the classroom. Thus, the teenagers are able to see how their programmed patterns are directly embroidered on T-shirts and bags. In future, the pattern-files can be sent via mail and the embroidered products can be picked up or shipped.

---

[9] As of 02.02.2018.

[10] https://share.catrob.at/luna/

[11] https://www.turtlestitch.org/

The project's outcomes will be 1) an extention of Luna&Cat, which includes the possibility to stitch pattern, 2) a gender-equitable framework for stitch/coding courses, and 3) to provide insights into the practical implementation for shops. As a result, not only app the app, but also gender-appropriate "guidelines" will be develop that show how young women could be motivated and, thus, they can serve as a guideline for others (teachers, trainers, schools, etc.). Figure 6 shows expressions from one of the first Stitching-Workshops.

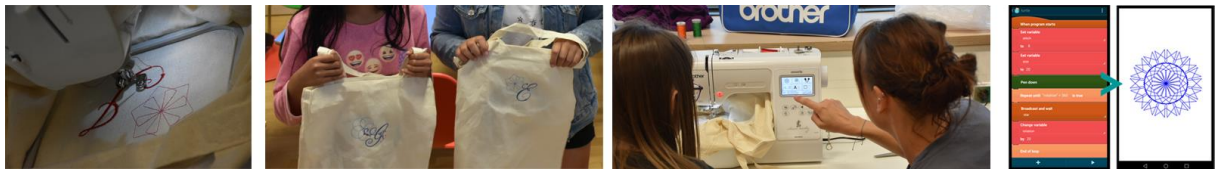

Figure 6: First expressions from a Stitching Workshop as part of the "Girls Coding Week 2018" at TU Graz.

## Acknowledgements

This work has been partially funded by Talente: FEMtech Forschungsprojekte 2017; Code'n'Stitch, eCall/FFG-Nr. 14975187 / 866755

---

[12] www.bits4kids.at

[13] www.apflbutzn.at